\documentstyle[12pt]{article}
\newcommand{\ie}{{\it i.e.}}
\newcommand{\eg}{{\it e.g.}}

\newcommand{\cf}{{\it cf.}}
\def\npb#1#2#3{{\it Nucl.\ Phys.} {\bf B#1} (19#2) #3}
\def\ijmp#1#2#3{{\it Int.\ J. Mod.\ Phys.} {\bf A#1} (19#2) #3}
\def\andjournal#1#2#3{{\bf #1} (19#2) #3}
\def\plb#1#2#3{{\it Phys.\ Lett.} {\bf B#1} (19#2) #3}
\def\prl#1#2#3{{\it Phys.\ Rev.\ Lett.} {\bf #1} (19#2) #3}
\def\mpla#1#2#3{{\it Mod.\ Phys.\ Lett.} {\bf A#1} (19#2) #3}
\def\cmp#1#2#3{{\it Commun.\ Math.\ Phys.} {\bf #1} (19#2) #3}
\def\prd#1#2#3{{\it Phys.\ Rev.} {\bf D#1} (19#2) #3}
\newcommand{\cpcp}{${\bf C}P^1\times {\bf C}P^1$}
\newcommand{\ztwo}{\mbox{${\bf Z}_2$}}
\renewcommand{\theequation}{\thesection \arabic{equation}}
\renewcommand{\thesection}{\arabic{section}.}
\renewcommand{\thesubsection}{\thesection \arabic{subsection}}
\renewcommand{\thefootnote}{\fnsymbol{footnote}}
\newcommand{\newsection}[1]{\vspace{1cm} \pagebreak[3]
  \addtocounter{section}{1} \setcounter{equation}{0}
  \setcounter{subsection}{0} \setcounter{footnote}{0}
  \begin{center} {\Large\rm \thesection\ #1} \end{center}
  \nopagebreak \medskip \nopagebreak}
\newcommand{\newsubsection}[1]{\vspace{7mm} \pagebreak[3]
  \addtocounter{subsection}{1} \noindent {\large\sc \thesubsection\ #1}
  \nopagebreak \vspace{5mm} \nopagebreak}
  \newcounter{app}
\newcommand{\oneappendix}[1]{\vspace{1cm} \pagebreak[3]
  \addtocounter{app}{1} \setcounter{equation}{0}
  \renewcommand{\theequation}{\Alph{app}.\arabic{equation}}
  \setcounter{footnote}{0} \begin{center} {\Large\rm Appendix:
  \\*[3mm] \Large\rm #1} \end{center} \nopagebreak
  \medskip \nopagebreak}
\def\undertext#1{\vtop{\hbox{#1}\kern 1pt \hrule}} %% USED FOR SUBSUBSECTIONS
\begin{document}
%%%%%%%%%%%%%%%%%%%%%%%%%%%%%%%%%%%%%%%%%%%%%%%%%%%%%%%%%%%%%%%%%%%%%%%%%%%%%
%  TITLE
\vglue1cm
\begin{center}
{\LARGE\rm Equivariant Topological Sigma Models}
\end{center}
\vspace{1cm}
%
%%%%%%%%%%%%%%%%%%%%%%%%%%%%%%%%%%%%%%%%%%%%%%%%%%%%%%%%%%%%%%%%%%%%%%%%%%%%%
%  AUTHOR
\begin{center}
{\large\rm Petr Ho\v{r}ava}
\end{center}
%%%%%%%%%%%%%%%%%%%%%%%%%%%%%%%%%%%%%%%%%%%%%%%%%%%%%%%%%%%%%%%%%%%%%%%%%%%%%
%  ADDRESS
\begin{center}
{\small\it Institute of Physics, Czechoslovak Academy of Sciences, \\
\small\it CS-18040 Prague 8, Czechoslovakia \\
\small\it and \\
\small\it Enrico Fermi Institute, University of Chicago,\footnote{Address
since September 1991.  E-mail: horava@yukawa.uchicago.edu} \\
\small\it Chicago IL 60637, USA}
\end{center}
\vspace{1cm}
%
%%%%%%%%%%%%%%%%%%%%%%%%%%%%%%%%%%%%%%%%%%%%%%%%%%%%%%%%%%%%%%%%%%%%%%%%%%%%%
%  ABSTRACT
\begin{center}
{\large Abstract}
\end{center}
\noindent
We identify and examine a generalization of topological sigma models suitable
for coupling to topological open strings.  The targets are K\"{a}hler
manifolds with a real structure, \ie\ with an involution acting as a complex
conjugation, compatible with the K\"{a}hler metric.  These models satisfy
axioms of what might be called ``equivariant topological quantum field
theory,'' generalizing the axioms of topological field theory as given by
Atiyah.  Observables of the equivariant topological sigma models correspond to
cohomological classes in an equivariant cohomology theory of the targets.
Their correlation functions can be computed, leading to intersection theory
on instanton moduli spaces with a natural real structure.  An equivariant
\cpcp\ model is discussed in detail, and solved explicitly.  Finally, we
discuss the equivariant formulation of topological gravity on surfaces of
unoriented open and closed string theory, and find a \ztwo\ anomaly explaining
some problems with the formulation of topological open string theory.
\vfill \baselineskip 16pt
\renewcommand{\thefootnote}{\arabic{footnote}}
\setcounter{footnote}{0}
\newpage
%  END OF TITLE PAGE
%%%%%%%%%%%%%%%%%%%%%%%%%%%%%%%%%%%%%%%%%%%%%%%%%%%%%%%%%%%%%%%%%%%%%%%%%%%%%
%
\newsection{Introduction}
\setcounter{equation}{0}
Witten's topological quantum field theory \cite{W:tqft,W:tsm,W:Trieste} has
convinced us recently of its intrinsic power. In particular, and quite
surprisingly, topological gravity in two dimensions has been shown to be
equivalent to the one matrix model \cite{W:gr,VV,W:Harvard}. On the other
hand, matrix models \cite{mm} has been invented in string physics as a
non-perturbative definition of string theory (working at least in toy
dimensions). Due to the intimate relations of the matrix models to the
Liouville theory, the equivalence of the topological two-dimensional gravity
to its ``physical'' counterpart is an amazing example of possible tight
relations between the conventional and topological versions of a physical
theory.

Two dimensional topological gravity has been completely solved (perturbatively
in the string coupling constant) by finding a set of recursion relations for
the correlation functions \cite{VV}. Having analyzed pure gravity, it is
natural to examine the structure of two-dimensional quantum gravity by looking
for natural and still sufficiently manageable generalizations of the simplest
case. Both of the above-mentioned approaches to two-dimensional quantum
gravity are known to offer such natural generalizations of their own. The
pure-gravity critical point of the one matrix model has found its natural
generalizations in the multi-critical points of the model and in the
multi-matrix models \cite{KDmmm}. In the language of conventional string
theory, these models describe strings in non-zero dimensions. From the
two-dimensional viewpoint they correspond to two-dimensional gravity coupled
to matter.

The topological approach also offers its own natural class of matter systems
that can be coupled to pure topological gravity without spoiling its
underlying symmetry, the so-called topological matter systems
\cite{W:gr,W:Harvard,W:surp}. One particular class of topological matter with
a direct geometrical interpretation are topological sigma models \cite{W:tsm};
coupling of topological sigma models to topological gravity describes
topological strings moving on a topological target manifold
\cite{W:gr,W:Harvard,topostring,DW}.

These two directions of generalizing the pure gravity system are closely
related to each other. In particular, the multi-matrix models have been argued
\cite{DW} to correspond to topological gravity coupled to a topological matter
system. This point has been further elaborated in \cite{Li}, where the matter
system has been identified with the topological version of $N=2$ mimimal
models, and the complete set of recursion relations for the correlation
functions of the theory has been found.

While the above-mentioned generalizations of the pure gravity system have been
thoroughly discussed recently, and much of their structure is clarified by
now, there is one interesting and manageable generalization of pure
topological gravity that has not received much attention. Namely, having
constructed pure gravity on closed orientable surfaces, one particularly
natural and simple generalization that, hopefully, could reveal some
interesting features of the theory, is pure gravity on surfaces with
boundaries and crosscaps, or in other words, theory of topological open
strings. Within the topological framework, this point was first addressed by
Myers \cite{Myers}, and by Hughes and Montano \cite{HM}. The authors of
\cite{HM} have considered pure topological gravity on surfaces with
boundaries, and discussed recursion relations for the theory. In the
matrix-model approach, open strings have been analyzed by Kazakov in the
Veneziano limit and then by Kostov in the double scaling limit \cite{openmm}.
As for the non-orientability phenomena in matrix models, these have been
analyzed thoroughly by Myers and Periwal, Harris and Martinec, and Br\'{e}zin
and Neuberger \cite{nonormm}.

The purpose of this paper is to present an ``equivariant'' approach to
topological matter on manifolds with boundaries and crosscaps. The origin of
this approach can be traced back to orbifold theories of critical string
theory \cite{DHVW}, and, in particular, to the fact that open string vacua can
be obtained by applying a generalized orbifold procedure to left-right
symmetric closed string vacua \cite{open,wso}. The ideas of \cite{open,wso}
have been used in open string model building \cite{Sagn}, and allowed for the
study of target duality in open string compactifications \cite{wsod}.
Moreover, the world-sheet orbifold approach to open strings has also proved
useful in relating conformal field theory open strings to Chern-Simons-Witten
theory in three dimensions \cite{csw}.

Inspired by this, we might consider field theories invariant (in the orbifold
sense) under a discrete symmetry group, acting possibly both on the space-time
and the field manifold. We encounter analogous situations frequently in
geometry, where the theories which are required to respect an additional
structure of a group action are usually called ``equivariant'' (equivariant
cohomologies, equivariant $K$-theory \cite{KR} {\it etc.}). As it turns out
that the topological sigma models we aim to investigate represent examples of
a more general notion of equivariant topological field theory defined in the
Appendix below, it seems natural to call them ``equivariant topological sigma
models.''

In cohomological field theories \cite{W:Trieste}, of which the equivariant
topological sigma model is an example, the functional integral for the
correlation functions of physical observables leads naturally to the
intersection theory for (co)homologies of moduli spaces (of instantons,
complex curves, or whatever corresponds to relevant classical configurations
of the particular Lagrangian). The equivariant approach to open strings has
the advantage that it can be easily combined with the cohomological approach
to the correlation functions of the theory, leading naturally to an
equivariant cohomology theory on the moduli spaces, thus making the theory a
relatively straightforward generalization of the non-equivariant case.

In fact, what we will really encounter in equivariant topological sigma models
is a very special case of equivariant theory. Our targets are complex
(K\"{a}hler) manifolds, with the orbifold group ${\bf Z}_2$ acting as a
complex conjugation on them. Such an antiholomorphic involution does not exist
necessarily on any manifold $M$, and defines what is called in algebraic
geometry ``a real structure'' on $M$ (see \eg\ \cite{Reel,real} for real
algebraic geometry). The set of fixed points of the involution is called
``the real part'' of $M$, and denoted by ${\bf R}M$. Analogously, the set of
all points of $M$ is denoted by ${\bf C}M$. (We will use this terminology and
notation throughout this paper.) The couple represented by a complex manifold
and an antiholomorphic involution on it then represents the
algebro-geometrical definition of a real manifold. Trying to construct
equivariant topological sigma models suitable for coupling to open strings, we
will thus be directed to the realm of real varieties in the sense just
mentioned. This is not too surprising, because surfaces with boundaries and
crosscaps that emerge in open string theory are themselves typical examples of
such real varieties \cite{AllGr,BCDCD}, and open string theory thus seems to
be just a real version of (complex) closed string theory.

Heuristically, we can expect a connection between topological open strings
and real algebraic geometry on purely mathematical grounds.  Indeed, one of
the deep results of topological sigma models is the fact that they offer
exotic topological invariants of pseudo-holomorphic curves in symplectic
manifolds as studied by Gromov and Floer \cite{W:gr,Floer,flo,gro}.
Mathematicians might then naturally ask: Is there a physical theory that gives
invariants for anti-holomorphic involutions of K\"{a}hler manifolds?  For a
K\"{a}hler target, recalling the form of the Lagrangians of both the
topological and the conventional sigma models, we can easily see that while
holomorphic diffeomorphisms preserving the K\"{a}hler structure are symmetries
of both of the Lagrangians, anti-holomorphic involutions of $M$ are not; they
do not represent symmetries of the topological sigma model Lagrangian even if
they preserve the K\"{a}hler structure. Indeed, the gauge-fixed topological
Lagrangian is dominated by instantons in the semiclassical limit, and
complex conjugations on $M$ map instantons to anti-instantons. To save the
day, we have to supplement the target complex conjugation with another
operation that also changes instantons to anti-instantons. A natural way to
do it is to invert the complex structure on the world-sheet, a procedure that
is known to lead immediately to open strings in the framework of world-sheet
orbifolds. Hence, we expect invariants associated to complex conjugations of
K\"{a}hler manifolds along the lines of \cite{W:tsm,W:gr} to be related to
topological open string theory.  This might represent one more reason why to
study equivariant topological sigma models.

This paper is organized as follows. Before studying two-dimensional
equivariant sigma models, we discuss general equivariant topological quantum
field theory (Section 2), and define it axiomatically in the Appendix. As a
particular example of such an equivariant theory, we examine
${\bf Z}_2$-equivariant topological matter in two dimensions. These models can
be considered as possible backgrounds for topological open strings. Most of
the general structure of the models can be inferred from the axioms
themselves, irrepective of a concrete realization of the matter system. In
particular, physical states that correspond to integrating out a boundary or
crosscap can be defined, representing a very simple, topological sigma model
version of the boundary and crosscap states well known in (conformal field
theory of) open strings \cite{bc}.

In Section 3, we specialize our discussion to (${\bf Z}_2$-) equivariant
topological sigma models with a K\"{a}hler target $M$. Observables are shown
to be in correspondence with the cohomological classes of an equivariant
cohomology theory on $M$. Functional integral representation of their
correlation functions is reduced by standard arguments to integrals over the
moduli spaces of (equivariant) instantons of the sigma model, and
consequently, to intersection theory in equivariant cohomologies. Some
particular examples (a \cpcp\ model, $K3$ surfaces) are studied in detail.

In Section 4 we examine the coupling of equivariant topological sigma models
to equivariant topological gravity representing the theory of topological open
strings.  First of all, we show that the boundary analogue of the puncture
operator does not have any descendants, a fact that can be understood either
on physical grounds or by using arguments from real algebraic geometry.  More
importantly, however, we demonstrate that the \ztwo\ symmetry to be gauged in
topological gravity is plagued by an anomaly, which prevents us from
constructing a non-trivial topological gravity on surfaces with boundaries
and crosscaps, at least within the equivariant framework adopted throughout
the paper.
%
%%%%%%%%%%%%%%%%%%%%%%%%%%%%%%%%%%%%%%%%%%%%%%%%%%%%%%%%%%%%%%%%%%%%%%%%%%%%
%
\newsection{Equivariant Topological Field Theory}
\setcounter{equation}{0}
A particularly natural way of thinking about topological field theory is the
axiomatic approach proposed by Atiyah \cite{axioms}, following the
axiomatization of $2D$ conformal field theory given by Segal \cite{Segal}. We
will closely follow reference \cite{axioms} in our treatment here (see also
\cite{DW:csw}).

In the axiomatic approach, one considers the $D$-dimensional topological
quantum field theory as a set of rules that associate to any oriented
$(D-1)$-dimensional manifold $\Sigma$ a vector space ${\cal H}_{\Sigma}$ (the
space of quantum states of the canonical quantization on $\Sigma\times
{\bf R}$), and to any oriented $D$-dimensional manifold $Y$ with boundary
$\partial Y$ a vector $\Psi _Y$ from ${\cal H}_{\partial Y}$ (the ``vacuum
state'' corresponding to the functional integral on $Y$) \cite{axioms}, in a
way compatible with general physical requirements summarized in a set of
axioms. Instead of repeating these axioms in this paper here (see
\cite{axioms}), we present in the Appendix their equivariant generalization,
\ie\ their extension to the case when there is an action of a fixed (discrete)
group $G$, both on space-time manifolds and on their Hamiltonian slices, which
must be respected by the data that define the topological quantum field
theory. The non-equivariant theory of \cite{axioms} is a special case of this
more general axiomatics for $G=1$.

Although there seems to be an explicit example of an equivariant topological
field theory in three dimensions \cite{csw}, the main field of application and
the power of the axioms is still in two dimensions, as we are now going to see.

%%%%%%%%%%%%%%%%%%%%%%%%%%%%%%%%%%%%%%%%%%%%%%%%%%%%%%%%%%%%%%%%%%%%%%%%%%%%%%%

\newsubsection{\sc Equivariant Topological Matter in Two Dimensions}

Our main concern in this paper are equivariant field theories suitable for
coupling to topological open strings, \ie\ equivariant topological matter
systems in $D=2$. Open strings are known to be related to closed string theory
via the ${\bf Z}_2$ orbifold procedure \cite{open,wso} that acts on the
world-sheet of the string as world-sheet parity, this action being possibly
supplemented with an action on the target. Hence, we will specialize
henceforth to ${\bf Z}_2$ equivariant topological theory in two dimensions,
and the word ``equivariant'' will mean ``${\bf Z}_2$-equivariant'' from now
on. The orbifold group ${\bf Z}_2$ will act on world-sheets and their
Hamiltonian slices by reversing their orientation, and the axioms thus require
a slight, obvious modification (which we do not present explicitly here).
Modding out the ${\bf Z}_2$-surfaces by the orbifold group will result in
surfaces with boundaries and crosscaps, referred to as Klein surfaces.  We
will sometimes use the same symbol $\Sigma$ both for a Klein surface and for
its orientable double, and call the real part of the double ``the boundary''
of $\Sigma$, with the hope of not causing any confusion.

Possible Hamiltonian slices of the equivariant world-sheets are disjoint
unions of ${\bf Z}_2$-manifolds of two possible topological types,
corresponding to doubles of the closed and open strings. Thus, physical states
in any equivariant topological matter system are divided into two classes
according to their world-sheet topology. We will denote the sector of closed
string physical states as ${\cal H}$, and pick a basis ${\cal O}_1,\ldots
{\cal O}_N$ in it. On account of the topological symmetry, each element of
${\cal H}$ represents a point-like observable living in the interior of the
world-sheet. Analogously, we will pick a basis in the open string Hilbert
space $\widetilde{\cal H}$,%
\footnote{As a rule, the objects that correspond to the open sector will be
marked by $\widetilde{\ }$, to distinguish them from those of the closed
sector.}
say $\widetilde{\cal O}_{\widetilde{1}},\ldots \widetilde{\cal
O}_{\widetilde{N}}$, with the corresponding point-like observables living at
the world-sheet boundary. To make our life simple, we will restrict ourselves
throughout this paper to the topological matter systems with only bosonic
observables in both sectors. The general case could be considered analogously.

Many crucial properties of topological matter can be inferred simply from the
axioms themselves (\cf\ the Appendix), irrespective of any possible underlying
structure of the theory. In particular, the factorization axiom represents a
powerful tool allowing the theory to be solved in terms of a few elementary
building blocks. Most of the structure of the non-equivariant topological
matter systems is known to be encoded \cite{W:gr,W:Harvard} in the operator
product expansion (OPE henceforth) of its BRST invariant observables:
\begin{equation}
{\cal O}_\alpha \cdot {\cal O}_\beta =\sum _{\gamma}{c_{\alpha\beta}}^{\gamma}
{\cal O}_\gamma,
\label{OPE1}
\end{equation}
which is valid independently of the location of the punctures the observables
are inserted in, as a result of the topological BRST symmetry. The OPE algebra
is an associative, commutative algebra, which we will assume to have an
identity, ${\cal O}_1$. (We will sometimes write tacitly 1 for ${\cal O}_1$.)
Having known the OPE algebra, the only ingredience needed to calculate
completely all genus zero amplitudes is the metric on the space of
observables, given by the two point function on the sphere:
\begin{equation}
\eta_{\alpha\beta}=\left\langle {\cal O}_{\alpha}{\cal O}_{\beta}
\right\rangle_0.
\end{equation}
Indices in the closed sector are lowered and raised by $\eta_{\alpha\beta}$
and its inverse.

At genus $g$, we can compute the correlation functions in the non-equivariant
theory, using factorization:
\begin{equation}
\left\langle \ldots \right\rangle _g=\sum _{\alpha \beta}\left\langle\ldots
{\cal O}_{\alpha}\right\rangle_{g-1}\eta^{\alpha\beta}\left\langle
{\cal O}_{\beta}\right\rangle _1,
\end{equation}
and carrying out explicitly the functional integral over the torus with the
${\cal O}_{\beta}$ insertion. This functional integral is equivalent to a
physical operator $W$, located in the puncture:
\begin{equation}
W=\sum _{\alpha \beta}\eta ^{\alpha \beta} \left\langle {\cal O}_{\beta}
\right\rangle_1{\cal O}_{\alpha}=\sum_{\alpha\beta\gamma}\eta^{\alpha \beta}
c_{\beta\gamma}{}^{\gamma}{\cal O}_{\alpha}.
\label{handle}
\end{equation}

Genus $g$ correlation function are given by
\begin{equation}
\left\langle \ldots \right\rangle _g=\left\langle \ldots W^g\right\rangle _0,
\end{equation}
thus completing the solution of the theory in terms of $c_{\alpha\beta}{}^{
\gamma}$ and $\eta _{\alpha \beta}$ \cite{W:gr}.

In equivariant topological matter theory, we can calculate any correlation
function in terms of a few boulding blocks using similar methods as in the
non-equivariant case. The first information we need is the action of the
orbifold group \ztwo\ on the space of observables, which splits it to even and
odd subspaces. We will denote the matrices representing the \ztwo\ action on
the closed and open sector by $\Omega_{\alpha}{}^{\beta}$ and
${\widetilde{\Omega}}_{\widetilde{\alpha}}{}^{\widetilde{\beta}}$
respectively.

Two-point functions of open observables on the disc define a metric in the
open sector:
\begin{equation}
\widetilde{\eta}_{\widetilde{\alpha}\widetilde{\beta}}=\left\langle
\widetilde{\cal O}_{\widetilde{\alpha}}\;\widetilde{\cal O}_{\widetilde{\beta}}
\right\rangle _{\rm disc}.
\end{equation}
Furthermore, we will need mixed two point functions of one closed and one open
state:
\begin{equation}
\widehat{\eta}_{\widetilde{\alpha} \beta}=\left\langle
\widetilde{\cal O}_{\widetilde{\alpha}}\;
{\cal O}_{\beta}\right\rangle _{\rm disc}.
\end{equation}
Open sector indices will be lowered and raised by
$\widetilde{\eta}_{\widetilde{\alpha}\widetilde{\beta}}$.

The operator product expansion of two open string observables located at the
same component of the world-sheet boundary is again an open string observable
(compare figure (1)). Simple topological considerations show that operator
product of one closed and one open state should be equal to a sum over open
states only (\cf\ figures (2), (3)). Consequently, the OPE algebra has the
structure of a semi-direct product:%
\footnote{We will assume that, similarly as the closed sector, the open sector
contains its vacuum state, with the corresponding operator (say
$\widetilde{\cal O}_{\widetilde{1}}$) being an identity in the OPE algebra
restricted to $\widetilde{\cal H}$.}
\begin{eqnarray}
\widetilde{\cal O}_{\widetilde{\alpha}}\cdot \widetilde{\cal
O}_{\widetilde{\beta}}&=&\sum _{\widetilde{\gamma}}
\widetilde{c}_{\widetilde{\alpha}\widetilde{\beta}}{}^{\widetilde{\gamma}}
\widetilde{\cal O}_{\widetilde{\gamma}},
\label{OPE2} \\
{\cal O}_{\alpha}\cdot \widetilde{\cal O}_{\widetilde{\beta}}&=&
\sum_{\widetilde{\gamma}}d_{\alpha \widetilde{\beta}}{}^{\widetilde{\gamma}}
\widetilde{\cal O}_{\widetilde{\gamma}}
\label{OPE3}
\end{eqnarray}
for some coefficients $\widetilde{c}_{\widetilde{\alpha}\widetilde{\beta}}
{}^{\widetilde{\gamma}}, d_{\alpha \widetilde{\beta}}{}^{\widetilde{\gamma}}$.
These coefficients are related to three point functions on the disc via
\begin{eqnarray}
\widetilde{c}_{\widetilde{\alpha}\widetilde{\beta}}{}^{\widetilde{\gamma}}&=&
\widetilde{\eta}^{\widetilde{\gamma}\widetilde{\delta}}
\left\langle \widetilde{\cal O}_{\widetilde{\alpha}}\;
\widetilde{\cal O}_{\widetilde{\beta}}\; \widetilde{\cal
O}_{\widetilde{\delta}}\right\rangle _{\rm disc}, \nonumber \\
d_{\alpha \widetilde{\beta}}{}^{\widetilde{\gamma}}&=&
\widetilde{\eta}^{\widetilde{\gamma}\widetilde{\delta}}
\left\langle {\cal O}_{\alpha}\; \widetilde{\cal
O}_{\widetilde{\beta}}\; \widetilde{\cal
O}_{\widetilde{\delta}}\right\rangle _{\rm disc}. \nonumber
\end{eqnarray}
Note that the $d_{\alpha\widetilde{\beta}}{}^{\widetilde{\gamma}}$ are not
independent of the structure constants we have already defined.  Physically,
the process of one open and one closed string state approaching each other is
not elementary, and can be decomposed into two elementary processes, using
factorization as in figure (3). On account of this, we arrive at the following
identity:
\begin{equation}
d_{\alpha \widetilde{\beta}}{}^{\widetilde{\gamma}}=\widehat{\eta}_{\alpha
\widetilde{\sigma}}\; \widetilde{\eta}^{\widetilde{\sigma}\widetilde
{\rho}}\;\widetilde{c}_{\widetilde{\rho}\widetilde{\beta}}
{}^{\widetilde{\gamma}}.
\label{d}
\end{equation}
The OPE algebra (\ref{OPE1}), (\ref{OPE2}) and (\ref{OPE3}), together with the
two point functions, allow one to calculate any $n$-point function at the
lowest genus, \ie\ on the disc in case we have at least one open observable
inside the correlator.

Higher genus correlation functions can be computed using operators that
represent integrating out a boundary or a crosscap. These boundary and
crosscap operators,
\begin{eqnarray}
B&=&\sum _{\sigma}b^{\sigma}{\cal O}_{\sigma},
\label{b} \\
C&=&\sum _{\sigma}c^{\sigma}{\cal O}_{\sigma},
\label{c}
\end{eqnarray}
are closed string operators obtained by performing the functional integral in
the theory on the world-sheet of the topology of a disc or a real projective
plane, both with one hole cut out of the surface (see figure (4)). The
coefficients in (\ref{b}),(\ref{c}) can be inferred from the explicit
expression for the functional integral:
\begin{eqnarray}
b^{\sigma}&=& \sum _{\beta} \eta ^{\sigma \beta} \left\langle
{\cal O}_{\beta}\right\rangle _{\rm disc} = \sum _{\beta}
\eta ^{\sigma \beta} \widehat{\eta}_{\beta \widetilde{1}} \\
c^{\sigma}&=& \sum _{\beta} \eta ^{\sigma \beta} \left\langle
{\cal O}_{\beta}\right\rangle _{{\bf R}P^2}.
\end{eqnarray}
These operators represent amusingly simple analogues of the boundary and
crosscap states known from critical string theory \cite{bc}. Making use of
these operators, we can compute any correlation function on surface $\Sigma$
with $h$ handles, $b$ boundaries, and $c$ crosscaps:
\begin{equation}
\left\langle\ldots\right\rangle_{(h,b,c)}=\left\langle\ldots\cdot W^h\cdot
B^{b-1} \cdot C^c\right\rangle_{\rm disc},
\end{equation}
where we have assumed for simplicity that all of the open string observables
are located at the same component of $\partial\Sigma$.

Let us now notice that one important topological fact constrains significantly
the structure of any equivariant topological theory in two dimensions. Indeed,
it is well known that one handle and one crosscap on a surface are
topologically equivalent to three crosscaps on the same surface, irrespective
of the rest of the topology of the surface. We have assumed that the
equivariant topological matter systems satisfy the factorization conditions
of the amplitudes, a crucial axiom of the theory. As a result of the
topological identity mentioned above, any surface with at least one handle and
one crosscap can be factorized in many different ways. The consistency
conditions of these factorizations are expressed in terms of an identity for
the handle and crosscap operators:
\begin{equation}
C\cdot W=C^3.
\label{topoid}
\end{equation}
We will refer to this equation as the ``topological identity'' below.

We will now show that, as a result of the topological identity, the crosscap
operator $C$ cannot be invertible as an element of the OPE algebra on
${\cal H}$ in any topological matter system with nontrivial $\Omega$. Indeed,
were the crosscap operator invertible, we would obtain by multiplying both
sides of the topological identity (\ref{topoid}) by $C^{-1}$ that $C^2$,
\begin{equation}
C^2=\sum _{\alpha \beta \gamma} c^{\beta}\; c^{\gamma}\; c_{\beta \gamma}
{}^{\alpha}{\cal O}_{\alpha},
\label{C2}
\end{equation}
should be equal to $W$, equation (\ref{handle}). Making now use of the fact
that one-point functions on the Klein bottle can be computed in two different
ways that are to be equivalent to each other:
\begin{equation}
\left\langle {\cal O}_{\alpha}\right\rangle _{\rm Klein\; bottle}=\left\langle
{\cal O}_{\alpha}C^2\right\rangle = {\rm Tr}({\cal O}_{\alpha}\Omega ),
\end{equation}
we get
\begin{equation}
\sum _{\beta \gamma}c_{\alpha \beta \gamma}\; c^{\beta}\; c^{\gamma} =
\sum_{\delta \epsilon} {c_{\alpha\delta}}^{\epsilon}\Omega_{\epsilon}
{}^{\delta}.
\end{equation}
Substituting this equation to (\ref{C2}), we can see that $C^2=W$ if and only
if $\sum_{\beta}c_{\alpha\beta}{}^{\beta}=\sum_{\beta\gamma}c_{\alpha\beta}
{}^{\gamma}\Omega_\gamma{}^{\beta}$.  Setting $\alpha =1$, this leads to ${\rm
Tr}(1)={\rm Tr}(\Omega )$, and thus to $\Omega =1$. Consequently, the
assumption that $C$ is invertible entails triviality of the action of the
orbifold group on ${\cal H}$.

In a theory with nontrivial $\Omega$, ant thus with $C$ non-invertible, the
conclusion of the previous paragraph is avoided by obtaining a weaker
coordinate expression
\begin{equation}
\sum _{\alpha \gamma \sigma \rho}{c_{\alpha \beta}}^{\gamma}\;
c^{\alpha}\; {c_{\gamma \sigma}}^{\rho}\left(\delta
_{\rho}^{\sigma}-{\Omega _{\rho}}^{\sigma}\right) =0,
\end{equation}
which is equivalent to the topological identity, and which can well be valid
with a non-trivial $\Omega$.

Using the OPE algebra, any $n$-point function can be reduced to a correlation
function with at most one closed string state, and at most one open string
state {\it at each boundary component of the world-sheet}. To solve the theory
completely, it is useful to know the operator that corresponds to integrating
out one boundary component with an open string observable,
$\widetilde{\cal O}_{\widetilde{\alpha}}$, inserted at it. Using
factorization, we get
\begin{eqnarray}
B_{\widetilde{\alpha}}&=&\sum _{\beta \sigma}\eta ^{\sigma
\beta}\left\langle {\cal O}_{\beta}\widetilde{\cal
O}_{\widetilde{\alpha}}\right\rangle _{\rm disc}{\cal O}_{\sigma}
\nonumber \\
&=&\sum _{\beta \sigma}\eta ^{\sigma \beta}\; \widehat{\eta}_{\beta
\widetilde{\alpha}}\; {\cal O}_{\sigma}.
\label{ba}
\end{eqnarray}

Beside the various boundary and crosscap operators, there is another object
already known from the ancient dual models \cite{upsi}, and studied thoroughly
in string field theory \cite{upsisft}, that can be found to have a simple
analogue in topological field theory: the upsilon operator of open string --
closed string transitions. Indeed, it can be easily shown that the operator
\begin{equation}
\Upsilon _{\widetilde{\alpha}}{}^{\beta}=\sum_\gamma
\widehat{\eta}_{\widetilde{\alpha}\gamma}\; \eta ^{\gamma \beta},
\label{up}
\end{equation}
intertwining from $\widetilde{\cal H}$ to ${\cal H}$, does exactly what the
open string -- closed string transition operator is supposed to do in the
equivariant topological matter system. This operator corresponds to the
functional integral on the surface of figure (5). On the other hand, operator
$\overline{\Upsilon}_{\alpha}{}^{\widetilde{\beta}}$ representing the inverse
process is obtained by raising and lowering the indices of
$\Upsilon_{\widetilde{\alpha}}{}^{\beta}$.

The simplicity of $\Upsilon$ reflected in equation (\ref{up}) is one more
example of the fascinating simplicity of topological field theory, and allows
one to observe some simple facts about the models.  For example, using the
upsilon operator, equation (\ref{ba}) can be written as
\begin{equation}
B_{\widetilde{\alpha}}=\Upsilon_{\widetilde{\alpha}}{}^{\beta}{\cal O}_{\beta},
\label{uppp}
\end{equation}
which is topologically obvious. Moreover, one expects some other simple
identities to be valid. To mention at least one example, we have two
expressions for the correlation functions on the surface with one hole and one
handle,
\begin{equation}
\left\langle {\cal O}_{\alpha}\right\rangle _{(1,1,0)}={\rm Tr}(\; \Upsilon
\overline{\Upsilon}\; {\cal O}_{\alpha})=\left\langle {\cal O}_{\alpha}\cdot
W\cdot B\right\rangle _0,
\label{upid}
\end{equation}
where in the trace formula the hole of the surface is generated by opening the
world-sheet by $\overline{\Upsilon}$ and closing it again by $\Upsilon$. All
of these particularities can be proved explicitly for equivariant topological
sigma models, which we will now examine.
\pagebreak
%
%%%%%%%%%%%%%%%%%%%%%%%%%%%%%%%%%%%%%%%%%%%%%%%%%%%%%%%%%%%%%%%%%%%%%%%%%%%%%
%
\newsection{Equivariant Topological Sigma Models}
\setcounter{equation}{0}
The basic multiplet of the topological sigma model with a K\"{a}hler target
$M$ is the BRST multiplet of target coordinates $X^\mu, \mu =1, \ldots 2m$,
and their topological ghosts:
\begin{equation}
\begin{array}{rcl}
[ Q,X^\mu ] &=&\psi ^\mu, \\
\{ Q,\psi ^\mu \} &=&0.
\end{array}
\label{brst}
\end{equation}
There are no gauge symmetries to be fixed in the topological sigma model
beside the topological symmetry, hence the absence of secondary ghosts in this
BRST algebra. Observables are thus to be constructed in a different way than
in topological Yang-Mills theory; they are known to correspond to de~Rham
cohomology classes of the target \cite{W:tsm}.

The most general Lagrangian ${\cal L}_{\rm t}$ that respects the deformation
symmetry underlying the BRST algebra (\ref{brst}), is a topological invariant
\cite{tsm}. Upon choosing a gauge fixing condition and introducing the
corresponding antighost -- auxiliary field BRST multiplet, we can write the
Lagrangian in its gauge fixed version \cite{tsm}:
\begin{equation}
{\cal L}_0={\cal L}_{\rm t} + \{ Q,\int _{\Sigma}\Psi\}
\end{equation}
The conventional choice for the gauge fixing fermion leads, after solving the
equations of motion for the auxiliary field, to the topological sigma model
Lagrangian of \cite{W:tsm}:
\begin{equation}
{\cal L}_0=\int _{\Sigma}\left\{ \partial
\bar{X}^{\bar{J}}\bar{\partial}X^IG_{I\bar{J}} -
\frac{i}{2}\left( \rho _{\bar{z}}^{I}D_z\bar{\psi}^{\bar{J}}+
\rho _z^{\bar{J}}D_{\bar{z}}\psi
^I\right)G_{I\bar{J}}-\mbox{$\frac{1}{4}$}\psi
^I\bar{\psi}^{\bar{I}}\rho _{\bar{z}}^J\rho _z^{\bar{J}}\;
R_{I\bar{I}J\bar{J}}\right\},
\label{gf}
\end{equation}
where $\rho$ are the antighosts and $R$ is the curvature tensor of the
K\"{a}hler metric.

Assume now that we are given an antiholomorphic target involution $\Omega$
that preserves the K\"{a}hler structure on $M$. With the choice of the gauge
fixing fermion as in (\ref{gf}), the gauge fixed Lagrangian is invariant under
simultaneous conjugation of the target and the world-sheet. Hence, we can try
to construct an equivariant topological field theory satisfying the axioms of
the previous section, via modding out the topological sigma model by the
${\bf Z}_2$ acting simultaneously on the world-sheet and the target.

As any other equivariant topological matter system in two dimensions, the
equivariant topological sigma model can be indeed viewed upon as a theory of
matter on surfaces with boundaries and crosscaps. The equivariant origin of
the theory leads to natural boundary conditions imposed on $X^{\mu}$ at
$\partial \Sigma$. Insisting on the validity of the BRST algebra has the
effect of fixing the boundary conditions on topological ghosts $\psi$ as well.
This knowledge is sufficient for identifying observables in the equivariant
topological sigma model. Upon choosing coordinates $(X^{I},\bar{X}^{\bar{I}})$
on $M$ in which the involution act as a pairwise conjugation:
\begin{equation}
\begin{array}{rccc}
\Omega:&X^I\rightarrow \bar{X}^{\bar{I}}, &&
\psi ^I\rightarrow \bar{\psi}^{\bar{I}},
\end{array}
\end{equation}
we have at the boundary of $\Sigma$:
\begin{eqnarray}
X^I|_{\partial \Sigma}&=&\bar{X}^{\bar{I}}|_{\partial \Sigma}, \\
\psi ^I|_{\partial \Sigma}&=&\bar{\psi}^{\bar{I}}|_{\partial\Sigma}
\end{eqnarray}
for the allowed mappings of $\Sigma$ to $M$. Hence, in the equivariant
topological sigma model the boundary of the world-sheet is constrained to be
mapped into ${\bf R}M$ (see figure (6)), and the ghost field $\psi$ is at the
boundary restricted to be (co)tangent to ${\bf R}M$.  At $\partial \Sigma$ we
thus have the restricted topological BRST algebra
\begin{eqnarray}
[\widetilde{Q},\widetilde{X}^I] &=&\widetilde{\psi}^I, \\
\{ \widetilde{Q},\widetilde{\psi}^I\} &=&0,
\label{brsto}
\end{eqnarray}
The $m$-tuple $\widetilde{X}^I$ of {\it real}\, coordinates, defined by
$\widetilde{X}^I\equiv X^I|_{\partial \Sigma}$, forms a coordinate system on
${\bf R}M$. (We have again used $\widetilde{\ }$ to distinguish objects living
at the boundary from those living in the interior of the world-sheet.)

Vertex operators of open string states in critical open string theory live on
$\partial \Sigma$. Analogously, open string physical observables of the
equivariant topological sigma models correspond to BRST invariant operators
composed of the fields that have survived at the boundary. Given a
differential form $\widetilde{A}$ on ${\bf R}M$,
\begin{equation}
\widetilde{A}=\sum \widetilde{A}_{I_1\ldots I_n}(\widetilde{X})\,
d\widetilde{X}^{I_1}\wedge \ldots \wedge d\widetilde{X}^{I_n},
\end{equation}
we construct out of $\widetilde{A}$ a composite operator $\widetilde{\cal
O}_{\widetilde{A}}$,
\begin{equation}
\widetilde{\cal O}_{\widetilde{A}}=\sum \widetilde{A}_{I_1\ldots I_n}
(\widetilde{X})\,\widetilde{\psi}^{I_1}\ldots \widetilde{\psi}^{I_n},
\end{equation}
localized in a point at $\partial \Sigma$. The BRST commutator of this
observable can be inferred from (\ref{brsto}), leading to
\begin{equation}
\{ \widetilde{Q},\widetilde{\cal O}_{\widetilde{A}}\} =\widetilde{\cal
O}_{\widetilde{d}\widetilde{A}},
\label{qform}
\end{equation}
where $\widetilde{d}$ is the exterior derivative on ${\bf R}M$. Hence,
$\widetilde{\cal O}_{\widetilde{A}}$ is physical for $\widetilde{A}$ a closed
differential form on ${\bf R}M$. Reading (\ref{qform}) from the right to the
left, we find exact differential forms on ${\bf R}M$ to be BRST trivial.
Consequently, non-trivial physical observables in the open sector are given by
de~Rham cohomological classes on ${\bf R}M$. The whole set of physical
observables in the equivariant topological sigma model is thus a direct sum of
the cohomologies of ${\bf C}M$ and ${\bf R}M$ over the reals:
\begin{equation}
{\cal H}\oplus \widetilde{\cal H}=H^{\ast}({\bf C}M,{\bf R})\oplus H^{\ast}
({\bf R}M,{\bf R}).
\label{hp}
\end{equation}
For further use we will pick a basis, ${\cal O}_1, \ldots {\cal O}_{N},
\widetilde{\cal O}_{\widetilde{1}},\ldots \widetilde{\cal O}_{\widetilde{N}}$,
in ${\cal H}\oplus \widetilde{\cal H}$.

The list of physical observables is not exhausted by these zero-form
observables. Indeed, as usual in cohomological field theories
\cite{W:Trieste}, observables ${\cal O}_{\alpha}, \widetilde{\cal
O}_{\widetilde{\beta}}$ give rise to a hierarchy of descent equations for BRST
invariant observables, wit the corresponding top form serving as a possible
new term in the Lagrangian. While any zero-form observable ${\cal O}_{\alpha}$
of the closed sector can be associated naturally via the descent equations a
one-form and a two-form observable, any zero form observable localized at the
boundary of the world-sheet generates just a one-form living at the boundary:
\begin{equation}
\begin{array}{rclcrcl}
d{\cal O}_{\alpha}&=&\{ Q,{\cal O}_{\alpha (1)}\} , &&
\widetilde{d}\widetilde{\cal O}_{\widetilde{\alpha}}&=&\{
\widetilde{Q},\widetilde{\cal O}_{\widetilde{\alpha}(1)}\}
, \\
d{\cal O}_{\alpha (1)}&=&\{ Q,{\cal O}_{\alpha (2)}\} , &&
\widetilde{d}\widetilde{\cal O}_{\widetilde{\alpha}(1)}&=&0,
\\
d{\cal O}_{\alpha (2)}&=&0, &&&&
\end{array}
\end{equation}
allowing the Lagrangian to be generalized by adding the BRST invariant
top-forms to
\begin{equation}
{\cal L}={\cal L}_0+\sum _{\alpha}a_{\alpha}\int _{\Sigma}{\cal
O}_{\alpha (2)}+\sum _{\widetilde{\alpha}}
\widetilde{a}_{\widetilde{\alpha}} \int _{\partial \Sigma}\widetilde{\cal
O}_{\widetilde{\alpha}(1)}.
\label{Lgen}
\end{equation}
Here only the invariant observables should contribute. After the functional
integral for correlation functions is reduced to calculations on an instanton
moduli space, the couple $(Q,\widetilde{Q})$ represents a couple of
exterior-derivative operators on the moduli space and on its real part
respectively.

The space of observables carries a representation of the orbifold group. In
the closed sector, the action of the orbifold group is induced naturally from
the underlying geometry: Given a ${\bf Z}_2$-action on a manifold, this
induces a natural ${\bf Z}_2$ action on the cohomology ring of this manifold,
\begin{equation}
\Omega ^{\ast}:H_{\ast}({\bf C}M,{\bf R})\rightarrow H_{\ast}({\bf C}M,
{\bf R}),
\end{equation}
via the pull-back of the cohomology classes by $\Omega$. This involution (on
integer cohomologies) is an important invariant in the theory of real
algebraic varieties \cite{real}. The $\Omega^{\ast}$ acting on the de~Rham
cohomologies of ${\bf C}M$ preserves the natural metric given by the
intersection form (supposing the complex dimension of $M$ is even), and serves
as the action of the orbifold group on the sector of closed observables of the
model.

To avoid problems with topologically non-trivial closed string configurations,
targets are usually required \cite{W:Harvard} to be simply connected,
$\pi_1({\bf C}M)=1$. For analogous reasons we should forbid open string
configurations of nontrivial topology in $M$. As a result of the equivariant
boundary conditions, both ends of the open string are constrained to sit in
the real part of the target. Nontrivial open string topologies would thus
typically emerge for ${\bf R}M$ not being connected. Hence, we will limit
ourselves throughout to the real structures with $\pi _0({\bf R}M)=1$.

Under these assumptions, we will argue that all of the open string observables
of the equivariant topological sigma model are even under the orbifold group
(unlike their analogues in a general equivariant two-dimensional matter
theory).

Heuristically, our arguments are as follows. By virtue of $\pi_0({\bf R}M)=1$,
any open string configuration is topologically trivial. Hence, classical
ground states carry zero energy by Hodge-theoretical arguments
\cite{W:tqft,W:tsm}, and are pointlike in the target. The orbifold group
action reduces on these point-like configurations to the pure target action.
This target action preserves ${\bf R}M$ pointwise, and the only possible
nontriviality of the action of the orbifold group on the open sector reduces
to the action on the vacuum, $\widetilde{\Omega} |\widetilde{0}\rangle =\pm
|\widetilde{0}\rangle$. This sign can be fixed from the OPE algebra, and all
of the open states are even under $\widetilde{\Omega}$.

In the case of non-equivariant topological sigma models, the OPE algebra is
known to give a deformation of the cohomology ring of $M$. One may thus wonder
what the classical structure of ${\cal H}\oplus \widetilde{\cal H}$ is, of
which the quantum OPE algebra can be expected to be a deformation. Notice
first that there is a natural structure of an ${\cal H}$-module on ${\cal H}
\oplus \widetilde{\cal H}$. Indeed, to define a multiple of a cohomology class
$\widetilde{\omega}$ on ${\bf R}M$ by a cohomology class $\omega$ on
${\bf C}M$, we will pull back the cohomology class $\omega$ from ${\bf C}M$ to
${\bf R}M$, and make the wedge product with $\widetilde{\omega}$. Equipped
with this natural structure, the module (\ref{hp}) associated with the pair
represented by a manifold $M$ and an involution on it, can be proved to be a
(non-classical) equivariant cohomology ring in the sense of \cite{Bredon}.%
\footnote{Note the important point that this equivariant cohomology theory is
{\it not} the same as the $G$-equivariant cohomology theory based on the
classification spaces $BG$ of $G$ (see \eg\ \cite{DW:csw} for the
latter).  It is interesting to see a non-classical equivariant cohomology
arising naturally in a physical context.}
We expect this equivariant cohomology theory to be recovered in the classical
limit of the OPE algebra of the equivariant topological sigma model.

Let us now consider the correlation functions, in the theory with
$a_{\alpha}=\widetilde{a}_{\widetilde{\alpha}}=0$ in the Lagrangian
(\ref{Lgen}). The functional integral definition of the correlation
functions,
\begin{equation}
\left\langle \ldots \right\rangle =\int {\cal D}X{\cal D}\psi{\cal D}\chi
{\cal D}\lambda \ldots e^{-{\cal L}_0},
\end{equation}
gets reduced by standard arguments to what might serve as the topological
definition of the correlation functions. In more detail, the functional
integral is dominated in the semi-classical limit by instantons, and as a
result of the BRST invariance, the semi-classical result is exact. After the
bosonic and fermionic determinants are cancelled against each other, the
functional integral is reduced to a finite-dimensional integral over the space
of instantons. Remembering now the equivariant structure of the theory, it is
easy to show that relevant to this calculation are equivariant instantons,
\ie\ those holomorphic mappings from the world-sheet to the target that are
invariant under the simultaneous action of the orbifold group on the target
and the world-sheet.

Given an equivariant instanton $\Sigma \rightarrow M$, we get the formal
dimension of its component of the moduli space of equivariant instantons using
an equivariant (or $KR$-theoretical \cite{KR}) version of the index theorem.
The dimension is equal to the number of equivariant infinitesimal deformations
of the instanton, minus the number of equivariant obstructions for integrating
these deformations. Without any calculation, this dimension can be determined
by observing that the involution on $\Sigma$ and $M$ induces an involution on
the moduli space ${\cal S}$ of {\it all}\,  instantons on $\Sigma$ which acts
as a complex conjugation with respect to the natural complex structure of
${\cal S}$, and that the moduli space of equivariant instantons is precisely
the real part ${\bf R}{\cal S}$ of ${\cal S}$. Then, supposing ${\bf R}{\cal
S}$ is non-empty,
\begin{equation}
{\rm dim}_{\; {\bf R}}\; {\bf R}{\cal S}={\rm dim}_{\; {\bf C}}\; {\bf C}
{\cal S},
\end{equation}
which is a general formula valid for any complex manifold with real structure.

Using these facts, we can find a topological description of the correlation
functions, analogously as in the non-equivariant case \cite{W:gr,W:Harvard}.
Let
\begin{equation}
\Phi :\Sigma \times {\cal S}\rightarrow {\bf C}M
\end{equation}
denote the universal instanton. Recall that $\Phi$ maps a point $z$ in
$\Sigma$ and an instanton $f$ in ${\cal S}$ to $f(z)$. It immediately follows
that the real part of the universal instanton is mapped to the real part of
$M$:
\begin{equation}
\Phi :\partial \Sigma \times {\bf R}{\cal S} \rightarrow {\bf R}M.
\end{equation}
Using this fact it is obvious that, picking a point $x$ at $\partial \Sigma$,
we can associate with any cohomology class $\widetilde{\alpha}$ on ${\bf R}M$
a cohomology class on ${\bf R}{\cal S}$, by first restricting $\Phi$ to $\{
x\} \times {\bf R}{\cal S}\equiv {\bf R}{\cal S}$, and then making the
pullback of $\widetilde{\alpha}$ via this restricted mapping to ${\bf R}{\cal
S}$. This cohomological class $\Phi ^{\ast}\widetilde{\alpha}$ is unique for
$x$ varying within one component of $\partial \Sigma$. It is known in the
non-equivariant theory \cite{tsm,W:gr} that we can associate with any
cohomology class $\alpha$ on ${\bf C}M$ a cohomology class $\Phi^{\ast}\alpha$
on ${\bf C}{\cal S}$. This cohomology class can be pulled back to ${\bf R}
{\cal S}$ by the canonical embedding $\iota :{\bf R}{\cal S}\rightarrow {\bf
C}{\cal S}$.

With this recipe, the correlation functions of the equivariant topological
sigma model reduce to integrations of the cohomology classes on the moduli
spaces of (equivariant) instantons:
\begin{equation}
\left\langle {\cal O}_{\alpha_1}\ldots {\cal O}_{\alpha_{n}}\cdot
\widetilde{\cal O}_{\widetilde{\beta} _1}\ldots
\widetilde{\cal O}_{\widetilde{\beta}_{s}}\right\rangle =
\int _{{\bf R}{\cal S}}\iota ^{\ast}\Phi ^{\ast}\alpha _1\wedge
\ldots \wedge \iota ^{\ast}\Phi ^{\ast}\alpha _{n}\wedge \Phi
^{\ast} \widetilde{\beta} _1\wedge \ldots \wedge
\Phi ^{\ast}\widetilde{\beta}_{s} .
\end{equation}
This formula reduces in the classical limit, in which only the homotopically
trivial instantons contribute, to
\begin{equation}
\left\langle {\cal O}_{\alpha _1}\ldots {\cal O}_{\alpha _{n}}
\cdot \widetilde{\cal O}_{\widetilde{\beta} _1}\ldots
\widetilde{\cal O}_{\widetilde{\beta}_{s}}\right\rangle =
\int _{{\bf R}M}\iota ^{\ast}\alpha _1\wedge \ldots \wedge
\iota ^{\ast}\alpha _{n}\wedge \widetilde{\beta} _1\wedge
\ldots \wedge \widetilde{\beta}_s ,
\end{equation}
which corresponds precisely to the equivariant cohomology theory mentioned
above.

The cohomological expression for the correlation functions can be translated
to the dual language of intersection numbers of homology classes using
Poincar\'{e} duality \cite{Spanier}. Let us denote by $M_{\alpha}$ (resp.\ by
$\widetilde{M}_{\widetilde{\beta}}$) a representant of the Poincar\'{e} dual
of cohomology class $\alpha$ on ${\bf C}M$ (resp.\ $\widetilde{\beta}$ on
${\bf R}M$). The Poincar\'{e} duals of the cohomology classes $\iota^{\ast}
\Phi ^{\ast}\alpha$ and $\Phi ^{\ast}\widetilde{\beta}$ on the moduli spaces
are (homologous to) the set of those instantons that map the point of the
world-sheet in which the observable ${\cal O}_{\alpha}$ of $\widetilde{\cal
O}_{\widetilde{\beta}}$ is inserted, to $M_{\alpha}$ or
$\widetilde{M}_{\widetilde{\beta}}$ respectively. Denote these Poincar\'{e}
duals by $L_{\alpha}$ and $\widetilde{L}_{\widetilde{\beta}}$, and note that
the Poincar\'{e} dual of the pullback $\iota ^{\ast}\Phi^{\ast}\alpha$ is
homologous to the intersection of $L_{\alpha}$ with ${\bf R}{\cal S}$; we will
denote this homology class on ${\bf R}{\cal S}$ by $\widetilde{L}_{\alpha}$.

Using this machinery, we can write down the dual expression for the
correlation functions:
\begin{equation}
\left\langle {\cal O}_{\alpha _1}\ldots {\cal O}_{\alpha _{n}}
\cdot \widetilde{\cal O}_{\widetilde{\beta} _1}\ldots
\widetilde{\cal O}_{\widetilde{\beta}_{s}}\right\rangle =
\# \left( \widetilde{L}_{\alpha _1}\cap \ldots \cap
\widetilde{L}_{\alpha _n}\cap
\widetilde{L}_{\widetilde{\beta}_1}\cap \ldots \cap
\widetilde{L}_{\widetilde{\beta}_s}\right) ,
\end{equation}
where the intersection number is computed in ${\bf R}{\cal S}$.

For these rather formal expressions to have sense, we must ensure some obvious
facts. First, the Poincar\'{e} duality only works for oriented manifolds.
While the moduli spaces ${\bf C}{\cal S}$ are canonically oriented, their real
parts can in principle be non-orientable. Orientability of moduli spaces is
important, but in general hard to prove \cite{FreedU}. We will thus simply
assume that our moduli spaces are orientable, as will be the case in the
example we will consider below.

Second, aiming to mod out the non-equivariant theory by $\Omega$, we should
ensure that it is a symmetry of the theory, even at quantum level; the
zero-mode measure of the functional integral may be non-invariant under
$\Omega$ on some components of ${\cal S}$, and violate the $\Omega$ symmetry
by a \ztwo\ anomaly.  The requirement of absence of such anomalies will
restrict the dimension of the target to even complex-dimensional targets, and
an analogous \ztwo\ anomaly will return again in Section 4 where we consider
equivariant topological gravity.
\vspace{.5cm}

\undertext{More General Boundary Conditions}

Up to now we have studied the simplest type of boundary conditions, namely
those induced from a single \ztwo\ involution on the target.  In \cite{flo}
and \cite{gro}, Floer and Gromov studied more general classes of boundary
conditions on pseudo-holomorphic curves in symplectic manifolds with almost
complex structures.  While Floer considered curves with different boundary
components embedded into different Lagrangian submanifolds $L_i$ ($i=1,\ldots
,n$) in the target, Gromov studied the same situation with the $L_i$'s being
totally real submanifolds.  It would be desireable to have a
quantum-field-theoretical description of these cases.  Here we comment briefly
on what is required in order to obtain such a description.

As for the class of boundary conditions studied by Gromov \cite{gro}, they can
be easily incorporated into the equivariant setting as follows.  Let us first
find, for each $L_i$, an antiholomorphic involution $\Omega_i$ which has $L_i$
as its set of fixed points.  This can be done at least locally \cite{gro},
in the sense that for each $L_i$ there exists such an involution on a
neighborhood of $L_i$ in the target, provided the complex structure is
integrable and $L_i$ is real analytic in the target \cite{gro}.  (This
situation is generic enough, as any complex structure can be deformed near
$L_i$ to allow for such an involution \cite{gro}.  For physically interesting
cases, one can use BRST invariance to prove independence of physical results
on the deformation, see \cite{W:tqft,W:tsm} for an analogous situation.)
These $\Omega_i$'s generate a discrete group (denoted by $G$) of automorphisms
of the target.  $G$ can now be used as an orbifold group (or a group of
equivariance in the sense of the Appendix) on our theory.  Obviously, the
resulting orbifold model allows us to study the desired pseudo-holomorphic
curves with different boundary components in different Lagrangian submanifolds
$L_i$.  However, there is one important technical point that prevents us from
computing more.  Namely, the orbifold model can be looked upon as an
(orbifold) theory on $M/G_0$, where $G_0$ is the subgroup of holomorphic
automorphisms contained in $G$.  Obviously, the fundamental group of $M/G_0$
is $G_0$ (with $M$ simply connected, in accord with general assumptions of
this paper).  This brings us to the as yet mostly unexplored theory of
topological sigma models with targets $N$ of $\pi_1(N)\neq 1$.  Exactly these
technical reasons have led the author to the decision not to consider the more
general class of equivariant topological sigma models in this paper (\cf\ the
discussion that follows eq.\ (3.15)).

As for the class of boundary conditions studied by Floer \cite{flo}, the
situation is essentially similar.  Indeed, given a generic Lagrangian
submanifold in the target manifold $M$ and assuming that the symplectic
structure tames the almost complex structure on $M$ \cite{gro}, it can be
easily shown that the Lagrangian submanifold is totally real.  In this sense
we end up with the situation analyzed in the previous paragraph.  Still, it
does not seem that the fully general case can be incorporated easily into the
equivariant approach, the point being that we might fail to find, for a
generic Lagrangian submanifold, an involution defined globally on the whole
target.  Hence, we encounter orbifolds that are only locally modelled by
factors of domains in ${\bf R}^n$ by discrete groups, without being globally
factors of a manifold by a discrete group.  This would require a
generalization of the framework for equivariant theories as we present it in
this paper.

\pagebreak
%
%%%%%%%%%%%%%%%%%%%%%%%%%%%%%%%%%%%%%%%%%%%%%%%%%%%%%%%%%%%%%%%%%%%%%%%%%%%%

\newsubsection{\sc Examples}

The simplest example of a non-equivariant topological sigma model discussed in
\cite{W:gr} is the topological ${\bf C}P^1$ model. Nevertheless, ${\bf C}P^1$
is not suitable as a target for the equivariant topological sigma model.
Indeed, although the OPE algebra of the ${\bf C}P^1$ model is symmetric under
any complex conjugation of the target, the metric on the space of observables
(given by the intersection matrix) is not.  This is indeed an example of the
\ztwo\ anomaly mentioned in the previous section.

To avoid the \ztwo\ anomaly, we have to look for another target.  Since we
would like to illustrate the structure of the theory in the open sector, we
want the real part of the target to be cohomologically non-trivial. Typical
manifolds of nontrivial cohomology are spheres. It is easy to find a complex
manifold $M$ with a real structure such that the real part ${\bf
R}M$ is isomorphic to the $m$-sphere, $S^m$. Indeed, such a
manifold can be obtained by writting the equation for $S^m$ as a
submanifold in ${\bf R}P^{m+1}$, which in homogeneous real
coordinates $\zeta _i$, $i=1,\ldots m+2$ on ${\bf R}P^{m+1}$
reads:
\begin{equation}
\zeta _1^2+\ldots +\zeta _{m+1}^2-\zeta _{m+2}^2=0,
\end{equation}
and then taking this equation as defined over ${\bf C}$. This
defines a complex algebraic variety $N$ in ${\bf C}P^{m+1}$.
Clearly, the standard real structure $\zeta _i\rightarrow
\bar{\zeta}_i$ induces on it a real structure of which $S^m$ is
the real part.

For $m=2$, we get a quadric in ${\bf C}P^3$. As a result of the
uniqueness (up to isomorphism) of the non-degenerate symmetric
quadratic form on ${\bf C}^4$, any two smooth quadrics in ${\bf
C}$ are isomorphic to each other \cite{GH}, and consequently, to
\cpcp . The real structure that is induced on \cpcp\ from the
involution we started with, maps the two ${\bf C}P^1$ factors to
each other, just reversing their complex structure.

The equivariant topological sigma model with this target
will be our main example in this paper. Before going to analyze
its structure, let us notice that the construction of the target
offers at least two hints how to construct targets for
equivariant topological sigma models. First, any smooth algebraic
variety in ${\bf C}P^m$ with real coefficients defines a real
algebraic variety which may (or may not) carry a K\"{a}hler
structure. Second, given any complex manifold $M$ and denoting
by $\overline{M}$ the same topological manifold with the reversed
complex structure, the product $M\times \overline{M}$ carries a
natural real structure that maps the first factor of this product
to the second one. Note that the moduli spaces of equivariant
instantons are orientable in this class of models.
\vspace{.5cm}

\undertext{The Equivariant \cpcp\ Model}

Closed sector observables of the model are in one-to-one
correspondence with real cohomology classes of \cpcp\ \cite{W:gr}.
The cohomology groups of \cpcp\ are spanned by a constant
function $1$, a volume 2-form $\omega _1$ of the first
${\bf C}P^1$ factor, a volume 2-form $\omega _2$ of the
second ${\bf C}P^1$ factor, and their wedge product $\omega\equiv
\omega _1\wedge \omega _2$. The complex structure of \cpcp\
induces on both of the ${\bf C}P^1$ factors a natural
orientation. Our conventions are such that $\omega _1$ coincides
with the orientation of the first ${\bf C}P^1$, while $\omega _2$
is opposite to the natural orientation of the second ${\bf
C}P^1$. We will use symbols ${\cal O}_1,{\cal O}_{\omega _1}, {\cal
O}_{\omega _2},{\cal O}_{\omega}$ to denote the
local observables that correspond to these cohomology classes.
Moreover, it will be convenient to fix a coordinate system
$Z_1,Z_2$ on the target such that the involution $\Omega$ we aim
to study takes $Z_1$ to $\bar{Z}_2$ and $Z_2$ to $\bar{Z}_1$.

The intersection form on \cpcp\ defines the metric on the space
of observables, with the non-zero elements
\begin{equation}
\eta _{1\omega}=\eta _{\omega _1\omega _2}=1.
\label{bubu}
\end{equation}

To be able to study the equivariant version of this model, we
will first determine the genus zero correlation functions,
\begin{equation}
{\cal F}_{m_1m_2m}\equiv \left\langle \underbrace{{\cal O}_{\omega
_1}\ldots {\cal O}_{\omega _1}}_{m_1}\cdot \underbrace{{\cal
O}_{\omega _2}\ldots {\cal O}_{\omega _2}}_{m_2}\cdot
\underbrace{{\cal O}_{\omega}\ldots {\cal
O}_{\omega}}_{m}\right\rangle _0,
\end{equation}
of the non-equivariant \cpcp\ model. First, we have two
independent ghost numbers in the \cpcp\ model, corresponding to
the homotopy classes of mappings from the sphere to the two ${\bf
C}P^1$ factors of the target. Zero modes of the functional
integral must be absorbed by the net ghost number of the
observables in the correlator, which gives two conditions on the
dimension of the instanton moduli space that can contribute to
the correlation function. Instantons of the instanton number
$(k,\ell )$ are given by:
\begin{equation}
Z_1=a_1\frac{\prod _{i=1}^k(z-b_{1,i})}{\prod
_{i=1}^k(z-c_{1,i})}, \;\;\;\;\;
Z_2=a_2\frac{\prod _{i=1}^{\ell}(z-b_{2,i})}{\prod
_{i=1}^{\ell}(z-c_{2,i})}.
\label{inst}
\end{equation}
This component of the moduli space of instantons has (complex)
dimension $2k+2\ell +2$, as can be easily verified by direct
counting of the independent parameters in (\ref{inst}).Each of
the instanton numbers can be weighted in the correlation
functions by an independent ``coupling constant.'' We will denote
them $\beta _1$ and $\beta _2$.

Putting all this together, we get the following correlation functions:
\begin{equation}
{\cal F}_{m_1m_2m}=\left\{
\begin{array}{ll}
\beta _1^k\beta _2^\ell & \mbox{for $m_1+m=2k+1$ and $m_2+m=2\ell
+1$,} \\
0 & \mbox{otherwise.}
\end{array}
\right.
\end{equation}
Hence, the two point functions coincide with the intersection
numbers of the corresponding homology classes, and the three
point functions lead to the following OPE algebra:
\begin{eqnarray}
{\cal O}_1\cdot {\cal O}_{\rm anyth.}&=&{\cal O}_{\rm anyth.},
\nonumber \\
{\cal O}_{\omega _1}\cdot {\cal O}_{\omega _1}&=&\beta _1,
\nonumber \\
{\cal O}_{\omega _2}\cdot {\cal O}_{\omega _2}&=&\beta _2,
\nonumber \\
{\cal O}_{\omega _1}\cdot {\cal O}_{\omega _2}&=&{\cal O}_{\omega},
\\
{\cal O}_{\omega _1}\cdot {\cal O}_{\omega}&=&\beta _1{\cal
O}_{\omega _2}, \nonumber \\
{\cal O}_{\omega _2}\cdot {\cal O}_{\omega}&=&\beta _2{\cal
O}_{\omega _1}, \nonumber \\
{\cal O}_{\omega}\cdot {\cal O}_{\omega}&=&\beta_ 1\beta _2.
\nonumber
\end{eqnarray}
Thus, we can see that the OPE algebra of the \cpcp\ model is a
two-parameter deformation of the classical de~Rham cohomology
ring, to which the OPE reduces in the limit $\beta _1, \beta
_2\rightarrow 0$. In this classical limit, only the instantons
homotopic to a constant mapping contribute to the partition
functions. Moreover, we can see that the \cpcp\ theory is in the
obvious sense a product of two ${\bf C}P^1$ models.

Upon setting $\beta _1=\beta _2=\beta$, the quantum theory is invariant under
the complex conjugation that takes $Z_1\rightarrow \bar{Z}_2$. The complex
conjugation acts on the set of closed observables as
\begin{equation}
\begin{array}{rrclcrcl}
\Omega :&{\cal O}_1&\rightarrow &{\cal O}_1, &&{\cal O}_{\omega_1}&
\rightarrow &{\cal O}_{\omega _2}, \\
&{\cal O}_{\omega} &\rightarrow &{\cal O}_{\omega},&&{\cal O}_{\omega _2}
&\rightarrow &{\cal O}_{\omega _1}.
\end{array}
\end{equation}
We will now study the equivariant sigma model based on this
involution.

Observables in the open sector correspond to cohomology classes
of the real part of \cpcp , which is topologically ${\bf C}P^1$.
These are given by a constant function, $\widetilde{1}$, and a
volume form, $\widetilde{\omega}$. The corresponding observables,
$\widetilde{\cal O}_{\widetilde{1}}$ and $\widetilde{\cal
O}_{\widetilde{\omega}}$, are both
even under the action of the orbifold group.

Consider a surface $\Sigma$ with a fixed complex conjugation on
it. This conjugation defines a real structure on the moduli space
of all instantons on $\Sigma$, of which the space of equivariant
instantons is the real part. The involution on ${\bf C}{\cal S}$
takes instantons of instanton number $(k,\ell )$ to those of
instanton number $(\ell ,k)$. Consequently, the real part ${\bf
R}{\cal S}$ belongs entirely to the subspace with instanton
number $(k,k)$ for some $k$. In particular, computations of the
metric and OPE algebra will require to know the explicit form of
the instantons on the disc:
\begin{equation}
Z_1=a\cdot \frac{\prod _{i=1}^k(z-b_i)}{\prod
_{i=1}^k(z-c_i)}, \;\;\;\;\;
Z_2=\bar{a}\cdot \frac{\prod _{i=1}^{k}(z-\bar{b}_i)}{\prod
_{i=1}^{k}(z-\bar{c}_i)}.
\label{instr}
\end{equation}
These instantons are in one-to-one correspondence with the
instantons on the sphere in the non-equivariant ${\bf C}P^1$
model \cite{W:gr}.

The metric on the open sector is given by
\begin{equation}
\widetilde{\eta}_{\widetilde{1}\widetilde{\omega}}=1.
\end{equation}
(To be precise, we should distinguish here and in (\ref{bubu}) the
intersection form on the equivarant de~Rham cohomologies of \cpcp\
from the full metric on the space of observables as given by two
point functions. Indeed, despite the fact that the lowest, classical
contribution to these two point functions will be given by the
intersection form, we will see below that classical
contributions to some two point functions can acquire ``quantum
corrections'' from instantons of non-zero instanton number.)

The OPE in the open sector reads
\begin{eqnarray}
\widetilde{\cal O}_{\widetilde{1}}\cdot \widetilde{\cal O}_{\rm anyth.} &=&
\widetilde{\cal O}_{\rm anyth.}, \nonumber \\
\widetilde{\cal O}_{\widetilde{\omega}}\cdot \widetilde{\cal
O}_{\widetilde{\omega}} &=&
\widetilde{\beta}\widetilde{\cal O}_{\widetilde{1}}. \nonumber
\end{eqnarray}
We have denoted by $\widetilde{\beta}$ the ``coupling constant'' that
weights the contribution from the equivariant instantons of
instanton number $(k,k)$ by $\widetilde{\beta}^k$. It is easy to show,
making use of the factorization axiom, that $\widetilde{\beta}=\beta$.

The mixed two point functions can be computed analogously,
leading to
\begin{eqnarray}
&\widehat{\eta}_{1\widetilde{\omega}}=\widehat{\eta}_{\omega
_1\widetilde{1}}=\widehat{\eta}_{\omega _1\widetilde{1}}=1, & \nonumber \\
&\widehat{\eta}_{\omega \widetilde{\omega}}=\beta .
\label{omom}
\end{eqnarray}
Perhaps the only surprising result here may be the
$\beta$-dependence of the mixed two point function $\widehat{\eta}_{\omega
\widetilde{\omega}}$, and we will present here the computation
explicitly. It can serve as a typical example of calculations in
equivariant topological sigma models, and can thus be
illuminating.

To calculate the two-point function
\begin{equation}
\left\langle {\cal O}_{\omega}\widetilde{\cal
O}_{\widetilde{\omega}}\right\rangle _{\rm disc}
\label{twopoint}
\end{equation}
that defines this element of $\widehat{\eta}$, we must pull back both $\omega$
and $\widetilde{\omega}$ to the real part of the instanton moduli space via
the universal real instanton, and compute the integral of their wedge product
over ${\bf R}{\cal S}$. Or, in the dual language, we must count intersections
of the homology cycles that are Poincar\'{e} dual to these differential forms.
Although $\iota ^{\ast}\omega$ is zero simply because this is a 4-form on a
2-manifold, the pullback of $\omega$ to ${\bf R}{\cal S}$ via $\Phi \circ
\iota$ can be non-zero. On dimensional grounds, the only instanton number that
can give a non-zero contribution to (\ref{twopoint}), is $k=1$. In this
component of the moduli space of equivariant instantons, the Poincar\'{e} dual
of $\iota ^{\ast}\Phi ^{\ast}\omega$ is (homologous to) the submanifold
consisting of those instantons (\ref{instr}) with $k=1$ that map the generic
point of the world-sheet in which ${\cal O}_{\omega}$ is located, say $z=i$,
to the Poincar\'{e} dual of $\omega$ in \cpcp , say $(Z_1,Z_2)=(\infty ,i)$.
Analogously, the Poincar\'{e} dual of $\iota ^{\ast}\widetilde{\omega}$ is
(homologous to) the set of instantons that map $z=1$ to, say, $(Z_1,Z_2)=
(0,0)$. These two cohomology classes intersect in one point, thus giving
equation (\ref{omom}).

The information on the two point functions can be conveniently summarized as a
``metric'' on the space of all observables of the model:
\begin{equation}
H=\left( \begin{array}{cc}
\widetilde{\eta}_{\widetilde{\alpha}\widetilde{\beta}} &
\widehat{\eta}_{\widetilde{\alpha}\beta} \\
\widehat{\eta}_{\alpha \widetilde{\beta}} &
\eta _{\alpha \beta}
\end{array}
\right) =
\left( \begin{array}{cccccc}
0&1&0&1&1&0\\ 1&0&1&0&0&\beta \\ 0&1&0&0&0&1\\ 1&0&0&1&0&0\\
1&0&0&0&1&0\\ 0&\beta &1&0&0&0
\end{array}\right) .
\label{H}
\end{equation}
(We have used the quotation symbols here because $H$ can
degenerate. Actually, this ``metric'' degenerates for $\beta
=\frac{1}{4}$, which is, amusingly, the critical value of $\beta$
in the non-equivariant version of the model \cite{W:gr}. Indeed,
restoring the open string coupling constant $\lambda$ and
weighting the contributions to $H$ from surfaces of the Euler
characteristic $\chi$ by $\lambda^{-\chi}$, we get
\begin{equation}
\det H=4\lambda ^2\beta -1.
\label{det}
\end{equation}
On the other hand, the ``exact'' partition function \cite{W:gr} for
the non-equivariant \cpcp\ model, defined as the sum of the
partition functions over the Riemann surfaces of all genera
properly weighted by the closed string coupling constant $\lambda
_c$, is equal to
\begin{equation}
\langle 1\rangle _{\rm ``exact''}=\sum _g\lambda_{c}^{g-1}\langle
1\rangle _g=\frac{1}{\lambda _c}{\rm Tr}\left( W^{-1}\cdot
\frac{1}{1-\lambda _cW}\right) =\frac{4}{1-16\lambda _c^2\beta
^2}.
\label{ex}
\end{equation}
Comparing (\ref{det}) and (\ref{ex}) we easily observe that they
are singular for the same value of $\beta$, assuming the usual
relation between the closed string and open string coupling
constant dictated by factorization, $\lambda _c=\lambda ^2$, is
valid. It would be nice to know precise reasons for this
``coincidence'' of the singular values of $\beta$.)

The OPE algebra is completed by calculating the mixed terms:
\begin{eqnarray}
{\cal O}_1\cdot \widetilde{\cal O}_{\rm anyth.}&
=&\widetilde{\cal O}_{\rm anyth.},
\nonumber \\
{\cal O}_{\omega _1}\cdot
\widetilde{\cal O}_{\widetilde{\omega}}&=&\beta \widetilde{\cal
O}_{\widetilde{1}},\nonumber \\
{\cal O}_{\omega _2}\cdot
\widetilde{\cal O}_{\widetilde{\omega}}&=&\beta \widetilde{\cal
O}_{\widetilde{1}},\nonumber \\
{\cal O}_{\omega _1}\cdot \widetilde{\cal
O}_{\widetilde{1}}&=&\widetilde{\cal O}_{\widetilde{\omega}}, \\
{\cal O}_{\omega _2}\cdot \widetilde{\cal
O}_{\widetilde{1}}&=& \widetilde{\cal
O}_{\widetilde{\omega}}, \nonumber \\
{\cal O}_{\omega}\cdot \widetilde{\cal
O}_{\widetilde{1}}&=& \beta \widetilde{\cal
O}_{\widetilde{1}}, \nonumber \\
{\cal O}_{\omega}\cdot \widetilde{\cal
O}_{\omega}&=&\beta \widetilde{\cal
O}_{\widetilde{\omega}},\nonumber
\end{eqnarray}
which can be obtained either from (\ref{H}) using (\ref{d}), or by the direct
calculation of the corresponding three point functions.

The boundary state can be computed using $\left\langle {\cal
O}_{\alpha}\right\rangle _{\rm disc}=\left\langle
{\cal O}_{\alpha}\cdot B\right\rangle _0$, leading to:
\begin{equation}
B={\cal O}_{\omega _1}+{\cal O}_{\omega _2}.
\end{equation}
Quite analogously, the crosscap state can be computed,%
\footnote{Note the equality of the crosscap and boundary states, which is
reminiscent of a condition from modular geometry of open strings that requires
equality of the massless parts of the crosscap and boundary states of the
theory \cite{bc,csw}, necessary in order to get rid of BRST anomalies.}
\begin{equation}
C={\cal O}_{\omega _1}+{\cal O}_{\omega _2}.
\end{equation}

We can now compute the closed observables that correspond to
integrating out a boundary with an operator insertion on it.
Because $\widetilde{\cal O}_{\widetilde{1}}$ is invisible under
correlator, $B=B_{\widetilde{1}}$. The state that corresponds to
one $\widetilde{\omega}$-insertion at the boundary,
\begin{equation}
B_{\widetilde{\omega}}=\beta {\cal O}_1+{\cal O}_{\omega}.
\end{equation}
can be read off from $\left\langle {\cal O}_{\alpha}\widetilde{\cal
O}_{\widetilde{\omega}} \right\rangle _{\rm disc}=\left\langle {\cal
O}_{\alpha}\cdot B_{\widetilde{\omega}}\right\rangle _0$. Having
known $B$ and $B_{\widetilde{\omega}}$, the $\Upsilon$ operator
of open string -- closed string transitions can be inferred from
(\ref{uppp}), and identities such as (\ref{upid}) can be verified
by direct calculation.

Identifying now the state that corresponds to integrating out a
handle:
\begin{equation}
W=4{\cal O}_{\omega},
\end{equation}
we have identified all the necessary ingredients to analyze the
topological identity $C\cdot W=C^3$ in the \cpcp\ model.
The left hand side of equation (\ref{topoid}) reads
\begin{eqnarray}
C\cdot W&=&({\cal O}_{\omega _1}+{\cal O}_{\omega _2})\cdot
4{\cal O}_{\omega} \nonumber \\
&=&4\beta ({\cal O}_{\omega _1}+{\cal O}_{\omega _2}),
\end{eqnarray}
while the right hand side is
\begin{eqnarray}
C^3&=&({\cal O}_{\omega _1}+{\cal O}_{\omega _2})^3 \nonumber \\
&=&({\cal O}_{\omega _1}+{\cal O}_{\omega _2}) \cdot
2(\beta {\cal O}_1 +{\cal O}_{\omega}) \nonumber \\
&=&4\beta ({\cal O}_{\omega _1}+{\cal O}_{\omega _2}).
\end{eqnarray}
Hence, we have explicitly proved that the topological identity
\begin{equation}
C\cdot W=C^3
\end{equation}
is valid in the equivariant \cpcp\ model. This is desireable,
because otherwise the amplitudes would be in clash with
factorization.
\vspace{.5cm}

\undertext{${}_{}$Real $K3$ Surfaces}

Another important class of manifolds suitable as targets for a
topological sigma model are the $K3$ surfaces \cite{W:gr,DW}.
These are defined as complex surfaces (\ie\ they are of real
dimension four) with the vanishing first Chern class and the
first Betti number. Each $K3$ surface is a simply connected
K\"{a}hler manifold, carrying a Ricci flat metric. Hence, $K3$
surfaces are Calabi-Yau manifolds in the lowest non-trivial
dimension. For more details on the geometry of $K3$ surfaces, see
\eg\ \cite{K3}.

We now wonder whether the $K3$ topological sigma models can be
twisted by ${\bf Z}_2$ to produce an equivariant topological
sigma model. In particular, we must find out whether $K3$
surfaces can carry a real structure. Suppose first that
there exists a real structure on a $K3$ surface $X$. Its real
part ${\bf R}X$ is, if non-empty, of real dimension two. In this
paper we are only interested in orientable and connected ${\bf
R}X$ for reasons sketched above. Consequently, we are looking for
the real $K3$ surfaces with the real part an orientable,
connected Riemann surface.

The problem of existence of real structures on $K3$ surfaces has
recently been addressed in the literature \cite{Kh,real}, leading
to interesting results. While any two complex $K3$ surfaces are
topologically isomorphic, general results of \cite{real} show
that there are exactly 66 distinct topological types that can be
realized as the real part ${\bf R}X$ of a real $K3$ surface. All
of them are orientable \cite{Kh,real}. Discarding surfaces with
$\pi _0({\bf R}X)\neq 1$, 12 distinct topologies of ${\bf R}X$
still remain: the empty set, and surfaces with $g$ handles for
$g=0,\ldots 10$; see \cite{real}.

The space of physical observables of the theory is now generated
by the observables that correspond to the cohomology classes of
${\bf C}X$ and ${\bf R}X$. Denoting $X_g$ the real $K3$ surface
with the real part isomorphic to the Riemann surface of genus
$g$, a basis in the cohomology ring is given by a 0-form $1$, 22
two-forms $\omega _i, i=1,\ldots 22$, and a four-form $\omega$ on
${\bf C}X_g$ in the closed sector, whereas in the open sector we
have the basis consisting of a 0-form $\widetilde{1}$, $2g$
one-forms $\widetilde{\omega}_j, j=1,\ldots 2g$, and a two-form
$\widetilde{\omega}$ on ${\bf R}X_g$. The $2g$ observables
corresponding to the one-forms on ${\bf R}X_g$ behave as
fermions. We will thus stop our discussion of the equivariant
$K3$ models here, as we have decided to make our life simple, and
not to consider fermionic observables in this paper.
\pagebreak
%
%%%%%%%%%%%%%%%%%%%%%%%%%%%%%%%%%%%%%%%%%%%%%%%%%%%%%%%%%%%%%%%%%%%%%%%%%%%%
%
\newsection{Coupling to Equivariant Topological Gravity?}
\setcounter{equation}{0}
Equivariant topological matter systems we have defined and discussed in this
paper are natural candidates for coupling to two dimensional topological
gravity on surfaces with boundaries and crosscaps. In turn, it is natural to
expect that such a gravity theory would fit into the equivariant framework we
have advocated here, and could be conveniently formulated as an ``equivariant
topoplogical gravity.'' We will finish this paper by an attempt to construct
such an equivariant topological gravity. However, our discussion will be
rather tentative, as we have no definite answer to some crucial questions of
the equivariant theory.

Topological gravity can be described in many different ways. We will use the
formulation given by Verlinde and Verlinde in \cite{VV}. The basic multiplet
of the topological BRST symmetry is given by the spin connection and its ghost
partners:
\begin{eqnarray}
\{ Q,\omega \} &=&\psi _0, \nonumber \\
\{ Q,\psi _0 \} &=&d\gamma _0, \\
\{ Q,\gamma _0 \} &=&0. \nonumber
\end{eqnarray}
Bearing in mind the geometrical origin of $\omega$ and using the assumption of
BRST invariance, we fix the boundary conditions for this multiplet in the
equivariant theory. Under the orbifold group action, $\omega ,\psi_0$ and
$\gamma _0$ are odd. Consequently, $\gamma_0$ is zero at the boundary. This
action of $\Omega$ on the basic BRST multiplet can be uniquely extended to
the whole field content of the theory, compatibly with its symmetries.

Observables in the closed sector of the theory are given by the descendants
$\sigma _n$ of the puncture operator $P$:
\begin{eqnarray}
P&=&c\bar{c}\cdot \delta (\gamma )\delta (\bar{\gamma}), \\
\sigma _n&=&{\gamma _0}^n \cdot P.
\end{eqnarray}
The descendants, $\sigma _n$, are are even or odd under $\Omega$, depending on
$n$ being even or odd. In the open sector, we can define the boundary puncture
operator
\begin{equation}
\widetilde{P}=c\cdot \delta(\gamma ),
\end{equation}
but by virtue of the boundary conditions satisfied by $\gamma_0$, all
composites of the form ${\gamma _0}^n\widetilde{P}$ for $n\neq 0$ are zero! It
thus seems that the boundary puncture operator has no descendants.

There is a natural mathematical support to this conjecture. Analogously as in
the sigma model case, the moduli spaces of equivariant surfaces can be viewed
upon as real parts of the moduli spaces of Riemann surfaces \cite{Sepp}. The
Chern classes of the line bundles ${\cal L}_{(i)}$, which Witten has used in
the topological expression for the correlation functions, can be pulled back
to the moduli spaces of equivariant surfaces. It is thus natural to expect
that
\begin{equation}
\left\langle \sigma _{n_1} \ldots \sigma _{n_s}\cdot
\widetilde{P}\ldots \widetilde{P}\right\rangle =
\int _{{\bf R}{\cal M}}
\iota ^{\ast}c_1({\cal L}_{(1)})^{n_1}\wedge \ldots \wedge
\iota ^{\ast}c_1({\cal L}_{(s)})^{n_s}
\label{tt}
\end{equation}
might represent the (formal) topological expression for the correlation
functions on surfaces with boundaries and crosscaps, with possible insertions
of the boundary puncture operator at $\partial \Sigma$. (Indeed, the integral
is taken over the appropriate moduli space of equivariant surfaces with
punctures.) In this topological framework, boundary puncture descendants can
be expected to enter the topological expression for correlation functions by
inserting characteristic classes of a real line bundle on ${\bf R}{\cal M}$ on
the right hand side of (\ref{tt}). It is however a well known fact that
non-trivial characteristic classes of real line bundles take values in
${\bf Z}_2$-cohomologies \cite{char}, and their real versions are zero.%
\footnote{For the {\it cognoscenti}: the cohomology of the classifying space
$B{\bf Z}_2\equiv RP^\infty$ of real line bundles is given by
$H^0(B{\bf Z}_2,{\bf Z})={\bf Z},\ H^{2i}(B{\bf Z}_2,{\bf Z})=\ztwo$ for
$i\geq 1$, and zero otherwise.  After tensoring with the reals, the only
survivor is the generator of the zeroth cohomology, in which we recognize the
boundary puncture operator $\widetilde{P}$.}
This seems to offer some further support to the conjecture that there are no
descendants of the boundary puncture operator in equivariant topological
gravity. Note that problems with observables living on the world-sheet
boundary have also been observed by Myers in \cite{Myers}.

At the quantum level, the situation with equivariant topological
gravity is even worse than it might have appeared from our
classical considerations. Indeed, the correlation functions of
$\omega _n$ do not respect the ${\bf Z}_2$ symmetry that we would like to
use in the orbifold procedure. As a consequence of this \ztwo\ anomaly, we
cannot insist on the decoupling of $\Omega$-odd descendants from the
correlation functions, and the standard mechanism of constructing an
equivariant theory out of a non-equivariant one cannot be used.

One possibility of curing this problem may be connected with the doubling
phenomenon noticed in matrix models \cite{W:Harvard,double}.  It is well known
by now that topological gravity and Hermitian matrix models of even potential
are not naively equivalent; rather, the partition function of topological
gravity corresponds to the square root of the partition function of the matrix
model. This doubling of degrees of freedom might help in constructing an
equivariant version of topological gravity. Alternatively, we might hope to
extract some information from the conjectured topological definition of the
correlation functions on surfaces with boundaries and crosscaps, equation
(\ref{tt}).  Clearly, the structure of topological open string theory in the
gravitational sector remains still unclear and deserves further investigation.
\vspace{.5cm}

%%%%%%%%%%%%%%%%%%%%%%%%%%%%%%%%%%%%%%%%%%%%%%%%%%%%%%%%%%%%%%%%%%%%%%%%%%%%%%
\begin{center}
{\large\rm Note Added (September 1993)}
\end{center}
This paper was originally published as Prague Institute of Physics preprint
PRA-HEP-90/18 (December 1990); the present version contains just minor changes
and misprint corrections as compared to the original.

Since 1990, many exciting new ideas and results have indeed been obtained in
the subject of topological field theories in general, and in topological sigma
models in particular.  It is clearly impossible to review all of them in this
short note, and I will just mention those that seem to be most directly
connected to the present paper.

While the present paper discusses what is now called the {\bf A} model, it is
possible to formulate its mirror-related counterpart, the {\bf B} model (for
conformal targets), and study the still enigmatic mirror symmetry between
them \cite{mirrors}.  Also, for targets of complex dimension three, the sigma
model defines a legitimate (classical) string theory by itself
\cite{W:cswstring}, without any necessity of coupling to topological gravity
on the worldsheet.  For these relatively simple models, one can develop the
corresponding string field theory \cite{W:cswstring}, and observe that in the
open string sector the result coincides with the Chern-Simons gauge theory on
the real part of the target \cite{W:cswstring}.

Some important recent observations indicate \cite{holoano} that the structure
of topological sigma models is actually slightly more sophisticated than
believed thus far.  When considered carefully, the path integral of the
topological theory contains a holomorphic anomaly, which makes the ${\bf A}$
model background-dependent.  The topological sigma models considered thus far,
including those of the present paper, then represent a specific choice of the
backgound, defined by sending the corresponding coupling constants to infinity
(for details, see \cite{holoano,W:bi}).  The existence of holomorphic anomaly
seems to lead to some new insight into the puzzles of background independence
in string theory, at least in its toy-model topological incarnation
\cite{W:bi}.

The paper has been restricted to simply connected targets; recently, some
progress has been achieved in some simple cases with targets whose fundamental
group is non-trivial, such as the torus \cite{toptorus}.  The structure of
observables is much more complicated than for the simply-connected targets.
In the specific case of the target torus, the infinite fundamental group
generates an infinite number of physical states and an infinite-dimensional
spacetime symmetry algebra, leading to topological $w_\infty$ supersymmetry,
odd-symplectic geometry \cite{toptorus} and Batalin-Vilkovisky geometry
\cite{toptorus,getzler} in the target.  These results fit nicely into a broad
picture \cite{EFR} that indicates the existence of topological symmetry {\it
in the spacetime} of topological string theory.
\pagebreak
%
%%%%%%%%%%%%%%%%%%%%%%%%%%%%%%%%%%%%%%%%%%%%%%%%%%%%%%%%%%%%%%%%%%%%%%%%%%%%%

\oneappendix{Axiomatics of Equivariant TQFT}
Let $G$ be a discrete group, allowed to act effectively by
orientation-preserving diffeomorphisms on our ``spacetimes,'' making
$G$-manifolds out of them. To the same extend as Atiyah's axiomatics is
related to usual cobordisms of manifolds, the equivariant axiomatics will be
related to equivariant cobordisms of manifolds. To define a $G$-equivariant
topological quantum field theory, we first associate with every oriented
$(D-1)$-dimensional $G$-manifold $\Sigma$ a vector space ${\cal H}_{\Sigma}$
(of physical states). To any $D$-dimensional $G$-manifold $Y$ we assign an
element $\Psi _Y$ of ${\cal H}_{\partial Y}$, with the induced $G$-action on
$\partial Y$ implicitly understood, and assume that these data satisfy the
following system of axioms:
\begin{enumerate}
\item
{\bf Topological invariance.} An equivariant isomorphism $f:\Sigma\rightarrow
\Sigma '$ induces an equivariant isomorphism ${\cal H}(f):{\cal H}_{\Sigma}
\rightarrow {\cal H}_{\Sigma '}$ compatible with the $G$-action, and these
induced isomorphisms compose in the obvious way.
\end{enumerate}
Bearing in mind that the vector space associated to a given
$(D-1)$-dimensional manifold $\Sigma$ is the space of physical states on
$\Sigma$, it should behave appropriately under disjoint union of two
Hamiltonian slices:
\begin{enumerate}
\setcounter{enumi}{1}
\item
{\bf Multiplicativity.} If $\Sigma_1\cup\Sigma_2$ is the disjoint union, then
\begin{displaymath}
{\cal H}_{\Sigma_1\cup\Sigma_2}={\cal H}_{\Sigma_1}\otimes{\cal H}_{\Sigma_2},
\end{displaymath}
with the obvious $G$ action.
\end{enumerate}
The fact that transition amplitudes can be defined is ensured by
\begin{enumerate}
\setcounter{enumi}{2}
\item
{\bf Duality.} If $\Sigma ^{\ast}$ is $\Sigma$ with the opposite orientation,
then ${\cal H}_{\Sigma ^{\ast}}={{\cal H}_{\Sigma}}^{\ast}$ is the dual space.
\end{enumerate}
The central axiom of topological quantum field theory is the requirement of
associativity, or the factorization property of physical transition
amplitudes, which represents the possibility to sum over the full set of
intermediate states in any channel:
\begin{enumerate}
\setcounter{enumi}{3}
\item
{\bf Factorization of amplitudes.} If $\Sigma$ is a component of $Y$, and
$\Sigma^{\ast}$ is a component of $Y'$, then
\begin{displaymath}
\Psi _W=(\Psi _{Y'},\Psi _{Y})_{\Sigma},
\end{displaymath}
where we have denoted by $W$ the result of sewing $Y$ and $Y'$ along their
common boundary component $\Sigma$, and $(\ ,\ )_{\Sigma}$ denotes the
canonical pairing of ${\cal H}_{\Sigma}$ with ${\cal H}_{\Sigma ^{\ast}}$,
\ie\ the contraction in the corresponding indices. An analogous identity
should be valid for non-separating cuts as well.
\end{enumerate}
\begin{enumerate}
\setcounter{enumi}{4}
\item
{\bf Completeness.} The states assigned to manifolds $Y$ with boundary
$\partial Y=\Sigma$ span the whole vector space ${\cal H}_{\Sigma}$.
\end{enumerate}
If ${\cal H}_{\Sigma}$ have canonical identifications with their duals, as it
is the case when ${\cal H}_{\Sigma}$ are Hilbert spaces, it is natural to
require:
\begin{enumerate}
\setcounter{enumi}{5}
\item
{\bf Conjugation.} For any oriented $D$-dimensional manifold $Y$,
\begin{displaymath}
\Psi _Y=\Psi _{Y^{\ast}}.
\end{displaymath}
\end{enumerate}

We have assumed implicitly in these axioms that all of the physical states
are bosonic. The axioms could be obviously generalized to allow for fermions
as well.

One particular way of constructing equivariant topological field theory is to
mod out a topological field theory by the action of a discrete subgroup of its
symmetry group, which is a generalization of the orbifold construction known
from critical string theory. For this reason, $G$ is sometimes referred to in
the paper as the ``orbifold group.''
\newpage
%
%%%%%%%%%%%%%%%%%%%%%%%%%%%%%%%%%%%%%%%%%%%%%%%%%%%%%%%%%%%%%%%%%%%%%%%%%%%%%
%

%
%%%%%%%%%%%%%%%%%%%%%%%%%%%%%%%%%%%%%%%%%%%%%%%%%%%%%%%%%%%%%%%%%%%%%%%%%%%%
\vspace{.5cm}
\begin{center}
{\Large\rm Figure Captions}
\end{center}
\begin{enumerate}
\item
Stable open string geometries at the tree level. Inserting physical states in
the punctures of the discs, the functional integral will give
$\widehat{\eta}_{\alpha \widetilde{\beta}}$ and
$\widetilde{c}_{\widetilde{\alpha}\widetilde{\beta}\widetilde{\gamma}}$,
respectively.
\item
The disc geometry that gives the three point function $d_{\gamma
\widetilde{\alpha}\widetilde{\beta}}$ is reducible by factorization to two
stable discs with punctures, thereby leading to equation (\ref{d}).
\item
The operator product expansion of one closed and one open string observable
gives another open string observable, which has been ``evaluated'' in the
picture, using figure (2).
\item
Insertion of the boundary or crosscap state on the world-sheet corresponds to
integrating out a boundary or a crosscap.
\item
The surface that gives via the functional integral the $\Upsilon$ operator of
open string -- closed string transition.
\item
A typical open string configuration on an equivariant topological target
manifold. Both ends of the string are confined to the real part ${\bf R}M$ of
the target.
\end{enumerate}
%
%%%%%%%%%%%%%%%%%%%%%%%%%%%%%%%%%%%%%%%%%%%%%%%%%%%%%%%%%%%%%%%%%%%%%%%%%%%%%
%
\end{document}